# Tuning the perpendicular anisotropy of ferromagnetic films by thickness, width, and profile

G. Kopnov and A. Gerber*

Raymond and Beverly Sackler Faculty of Exact Sciences,

School of Physics and Astronomy, Tel Aviv University,

Ramat Aviv, 69978 Tel Aviv, Israel

Perpendicular magnetic anisotropy was found to be highly sensitive to the nominal thickness and morphology of laterally heterogeneous CoPd films in the vicinity of the metal – insulator transition. We used the effect to tune the anisotropy by the width of lithographically patterned stripes with non-uniform cross-sectional thickness profiles. The phenomenon and the method can be suitable to control the local anisotropy of spintronic logic circuitry elements by their shape and lateral dimensions.

* Corresponding author. Email: gerber@tauex.tau.ac.il

Perpendicular magnetic anisotropy (PMA) in thin magnetic films is a key feature in the design of spintronic devices. Magnetic tunnel junctions with perpendicular anisotropy form the basis of the spin-transfer torque magnetic random-access memory (STT MRAM), which is nonvolatile, fast, dense, and has quasi-infinite write endurance and low power consumption. The nonvolatile logic and computing devices built with the PMA materials are expected to replace or complement the semiconducting CMOS information technology [1]. In this context, tuning the local anisotropy is particularly important for the spintronic logic circuitry to control the nucleation and propagation of the magnetization reversal. Different techniques for enhancement or suppression of the local anisotropy have been suggested. Adding large end pads to nanowires increases the chance of finding a defect that can act as a reversal nucleation site [2]. However, a large area is needed to guarantee the presence of a defect which increases the element footprint and lowers the anisotropy. Control of the nucleation site can alternatively be obtained by localized Ga+ or He+ ion irradiation [3,4]. Tuning the magnetization reversal by in-plane shape and dimensions of the circuit elements has been suggested by Kimling et.al. [5] and Mansell et.al. [6] who found that the coercivity of nanowires decreased significantly as their width decreased below a certain critical value [6] or when the wire tip had triangularly pointed shape [5]. The ability to design controllable circuits by the planar geometry of the elements opens the possibility to fabricate the spintronic memory and logic devices in a single lithographic process.

The reported reduction of the coercivity with decreasing width [6] is unexpected for the perpendicularly magnetized ferromagnets. An opposite trend is usually observed in dots, nano-dots, and nano-wires. Enhancement of coercivity with decreasing dimensions has been attributed to the limitation of the number of the magnetization reversal nucleation sites [7, 8], enhanced domain wall pinning by edge defects [9-11] and a gradual transition from the domain wall pinning to the nucleation-dominant switching [12, 13]. An increase of the coercive field with decreasing width in FePt nanowires was attributed to the suppression of available paths for domain wall propagation [14]. For the same reason, notches are used in nanowire fabrication to trap domain walls [15]. On the other hand, reduction of the coercivity below a certain critical width was observed in several materials: Pt/CoFe/Pt tri-layers [16], Co/Pt and CoFeB/Pt multilayers [6], and antiferromagnetic $Mn_3Sn$ nanowires [17]. Similar behavior in different materials having different interfacial anisotropies, saturation magnetizations, exchange lengths, coercivity, and nucleation site density suggests the existence of a general trend. It was proposed [5] that reduction of the anisotropy in narrow patterned wires can be caused by shadowing effects due to the resist mask during sputter deposition. However, the conditions and the physical mechanism responsible for the suppression of the PMA have not been revealed.

We'll show that the origin of the PMA reduction in narrow wires can be related to their effective thickness rather than width. Anisotropy of laterally heterogeneous CoPd films in the vicinity of the metal–insulator transition was found to be highly sensitive to their nominal thickness and morphology. We used the effect to tune the coercivity by the width of lithographically patterned stripes with non-uniform cross-sectional profiles, where the effective thickness depended on the width.

Polycrystalline $Co_{20}Pd_{80}$ (atomic concentration) ferromagnetic films were fabricated by co-sputtering from separate Co and Pd targets on room temperature GaAs substrates at 5 mbar Ar pressure. The composition was controlled by the relative rf-power of the respective sputtering sources and tested by the energy dispersive x-ray spectroscopy analysis (EDAX). No post-deposition annealing was made. The first group of 3 nm to 100 nm thick samples was deposited through a square shape shadow mask with lateral dimensions $5 \times 5$ mm. The second group had the Hall bar shape, 5 to 100 μm wide and 3 nm to 100 nm thick, patterned using the photolithography with 1.4 μm thick AZ5214E photoresists. Resistance and Hall effect of the square-shape samples were measured using the Van der Pauw protocol. The Hall bars were measured in the standard 5-probe configuration. All measurements were done at 77K. Transmission electron microscopy (TEM) images were taken with JEOL JEM-2010F UHR device. The Hall bar cross-section profiles were scanned with the atomic force microscope (AFM) Park Systems NX10.

It was realized time ago that the extraordinary Hall effect resistance in ferromagnetic materials is an electric replica of magnetization, and measurement of the effect can be used for the study of ultrathin films [18, 19] and nano-scale objects [20, 21]. Fig.1 presents the normalized Hall resistance of a series of large square-shaped $Co_{20}Pd_{80}$ samples with different thicknesses as a function of normal to plane magnetic field. All samples exhibit strong perpendicular anisotropy with square hysteresis loops and full remanence (decreasing to 0.9 in the 6 nm thick sample). The coercive field as a function of thickness is shown in Fig.2a. $H_c$ is approximately constant in thick films, and drops sharply in a narrow thickness range below 10 nm. No hysteresis was observed in the 3 nm thick sample. A rapid drop in the coercivity is matched by a sharp rise of the resistivity, shown in Fig. 2b. Divergence of the resistivity at a critical thickness threshold is attributed to the laterally inhomogeneous morphology of ultra-thin films and is generally described by the percolation model [22, 23], as:

$$\rho(d) = \alpha \rho_0 (t - t_c)^{-m} \tag{1}$$

where $\alpha$ is the units transition coefficient, $\rho_0$ is the resistivity of the bulk solid with the same structure and composition as the film, $t$ is a nominal thickness, and $t_c$ is the critical thickness below which metallic clusters become finite and resistivity infinite. Exponent $m$ is 1.2 for 2-dimensional systems. The solid line in Fig. 2b is a fit of the thickness-dependent resistivity to Eq.1 with $t_c = 2.4 \pm 1$ nm and $m = 0.9 \pm 0.1$.

Transmission electron microscope (TEM) images of the nm-range thin films (Fig. 3) confirm their inhomogeneous meandrous morphology. Polycrystalline CoPd (dark) has the fcc structure with an average crystallite size of 2.3 nm in the 5 nm thick film. One can, therefore, connect the loss of the perpendicular anisotropy in thin films to the development of a strongly inhomogeneous percolation morphology. The critical coercivity thickness (5.9 nm) is higher than that of the resistivity (2.4 nm), meaning that the anisotropy is lost when the material is still interconnected and forms a long-range "infinite" cluster.

Sharp suppression of the coercivity with decreasing thickness in the vicinity of the percolation threshold contrasts the previous studies. Films with low nominal thickness are composed of disconnected ferromagnetic grains. If these particles are smaller than a certain critical size of the order of a domain wall width, the magnetization reversal proceeds through coherent rotation of the spins, and the coercive field is large. An increase in the number and size of large clusters allows a crossover from magnetization rotation to domain wall nucleation and motion that becomes the dominant magnetization switching mechanism. The decrease of the coercivity as the thickness is increased, correlated with the morphology change from the separated nano-particles to continuous layer has been observed in materials with both in-plane anisotropy [24 - 26] and also in the FePd L1$_0$ phase with perpendicular anisotropy [27, 28]. The transition can be non-monotonic which was explained [29] by the presence of empty sites in large clusters that pin the domain-wall displacement. In the limit of continuous films, the introduction of finely distributed nonmagnetic sites, acting as domain wall pinning centers, was suggested as a tool to increase the coercivity [30, 31]. Nano-patterned structures with ordered networks of nano-pores (anti-dots) have been fabricated with the in-plane anisotropy materials [32, 33], and the PMA multilayers and alloys [34, 35]. In all these studies the coercive field of the nanoporous films was higher than of the corresponding continuous films. Several mechanisms can be considered to understand the discrepancy. The perpendicular anisotropy in CoPd alloys is usually attributed to the magnetoelastic effect, which is a combination of tensile stress and a large negative magnetostriction constant [36-38]. Reduction of the anisotropy might be related to a release of tensile stress by voids. In addition, similar to Co/Pd and Co/Pt multilayers, the interfacial anisotropy can be affected by spatial dilution in the percolation range. On the other hand, the temperature at which the measurements were done (77 K) can be important. Finite-size clusters lose the anisotropy at their blocking temperature due to thermal fluctuations, and their number grows in the vicinity of the percolation threshold. Thus, the effect we observe can be the result of a subtle interplay between the magnitude of the anisotropy, size and distribution of finite clusters, and temperature.

Although we have no definite explanation of the anisotropy reduction in percolating $Co_{20}Pd_{80}$ films, one can adapt the effect to tune the anisotropy by the width. Standard fabrication of patterned films includes the optical or electron-beam lithography, sputter deposition, and lift-off processing of the resist. The resulting thickness profile is usually not uniform. Fig. 4 presents the typical thickness profiles of several Hall bar stripes with different widths. The profiles were scanned using an AFM microscope. The nominal thickness of this series was 20 nm. Shadowing of the incoming material flux by the 1.4 µm thick photoresist results in a gradual thickness increase from the edge inward. The range over which the thickness grows from zero to its predetermined value is 3 – 5 µm. As a result, 5 µm and 10 µm wide stripes have dome-shaped cross-sections, and their peak thickness doesn't reach the nominal value. Wider stripes have trapezoid-shaped profiles and the desired thickness in the middle. The base is 2.5 µm wider than the intended. The surface roughness of ± 2 nm corresponds to the crystalline size. Despite the imperfection of the fabrication, the resulting non-uniform profile is useful. The average thickness depends on the stripe width. The inhomogeneous percolating sections with the reduced anisotropy are located along the stripe's edges, and their density and coverage fraction depends on the profile and width.

The width effect on the hysteresis loop coercivity is illustrated in Fig.5. The normalized EHE resistance of several $Co_{20}Pd_{80}$ Hall bars with 10 nm nominal thickness and width between 5 μm and 500 μm are shown as a function of field. The general shape and remanence of the hysteresis don't change, while the coercivity decreases from 0.15 T in the 100 μm wide stripe to 0.06 T in the 5 μm wide one. The extent of the coercivity reduction and the critical width below which it occurs depend on the nominal thickness. The normalized coercive field as a function of the stripe width for bars with different nominal thicknesses is shown in Fig.6. The coercive field is normalized by its value in 100 μm wide stripes. The coercivity of thick films (20 nm and 50 nm) starts decreasing only in narrow stripes below 10 μm. In a thin 7 nm thick series, the reduction of coercivity starts already at 100 μm and drops by 80% in the 5 μm wide stripe.

The effects of thickness and width can be unified. Fig.7 presents the coercive field as a function of thickness for the large area square samples and as a function of the average thickness of narrow Hall bars. The latter was calculated from the measured cross-section profiles. All the data collapse on a single curve.

Our findings can be briefly summarized as follows: (i) the perpendicular anisotropy of $Co_{20}Pd_{80}$ films at relatively low (77 K) temperature drops to zero with decreasing thickness in a narrow range between 10 nm to 3 nm nominal thickness; (ii) the resistivity diverges in the same thickness range, and (iii) films have heterogeneous percolation morphology. We assume, therefore, that suppression of the perpendicular anisotropy is correlated with the laterally heterogeneous morphology and disappears at a certain "percolation" threshold. The second part relates to the correlation between the suppression of coercivity in narrow stripes and the cross-section profile of thin and narrow films. The cross-section of CoPd stripes fabricated by optical lithography is not uniform, with the thickness varying over a significant distance from the edge due to the resist mask shadowing. The fraction of the laterally heterogeneous section with the reduced coercivity depends on the nominal thickness, width, and profile of the film, which can be pre-designed and controlled. Gradual suppression of the anisotropy in narrow stripes correlated with the reduced average thickness has been demonstrated.

Identification of the PMA suppression mechanism in the topologically percolating range, and generality of the effect in other materials are open for further studies. For practical applications, the guidelines brought here have to be adapted to the materials with room temperature anisotropy and the nano-scale cross-sectional profiles.

The work was supported by the Israel Science Foundation grant No. 992/17.

The data that supports the findings of this study are available within the article.

# References.

**Figure captions.**

Fig.1. EHE resistance hysteresis loops of square-shape 5 mm x 5 mm samples with different thicknesses. T = 77 K.

Fig.2. Coercive field (a) and resistivity (b) of large area square-shape films as a function of thickness. Solid line in (b) is a fit to Eq.1 with $t_c = 2.4 \pm 1$ nm and $m = 0.9 \pm 0.1$.

Fig.3. TEM image of the 5 nm thick $Co_{20}Pd_{80}$ film deposited on carbon. CoPd is dark. The average fcc crystallite size is 2.3 nm.

Fig.4. AFM cross-section profiles (thickness as a function of distance from edge) of three stripes with 20 nm nominal thickness. The average thickness is 9.3 nm, 13.2 nm and 17 nm for the 5 µm, 10 µm, and 20 µm wide stripes respectively.

Fig.5. EHE hysteresis loops of 10 nm thick $Co_{20}Pd_{80}$ Hall bars with different width. T = 77K.

Fig.6. Normalized coercive field as a function of the stripe's width for different thicknesses.

Fig.7. Coercive field of large area square-shape samples (solid dots) and narrow Hall bars (open symbols) as a function the average thickness.

**Figures**

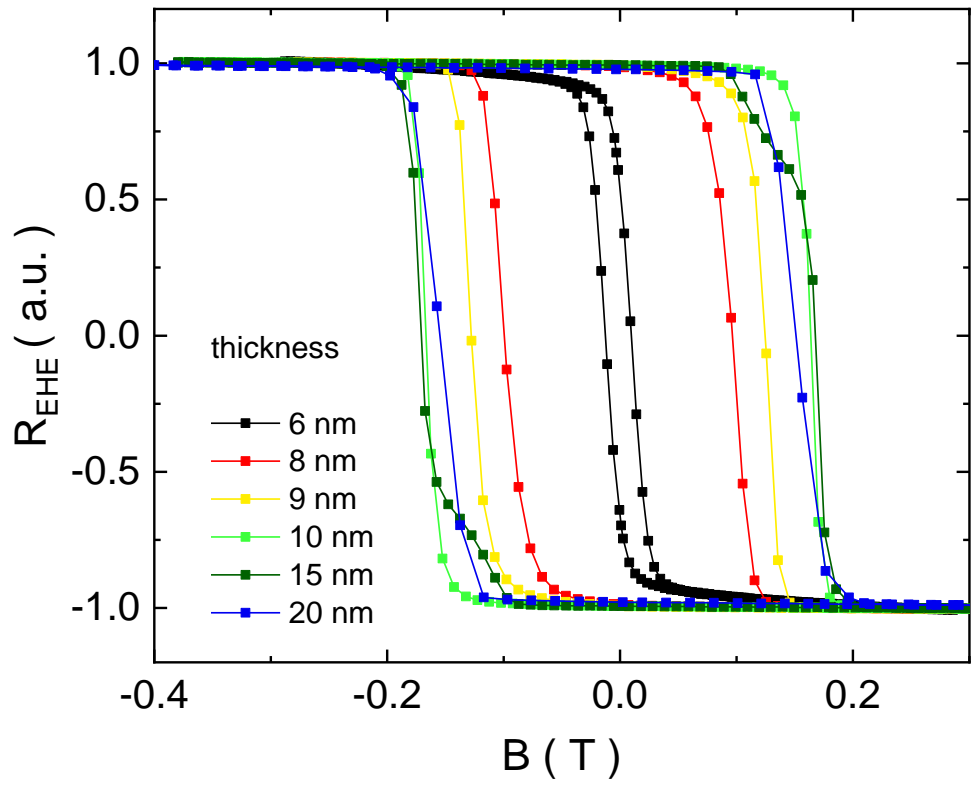

Fig. 1

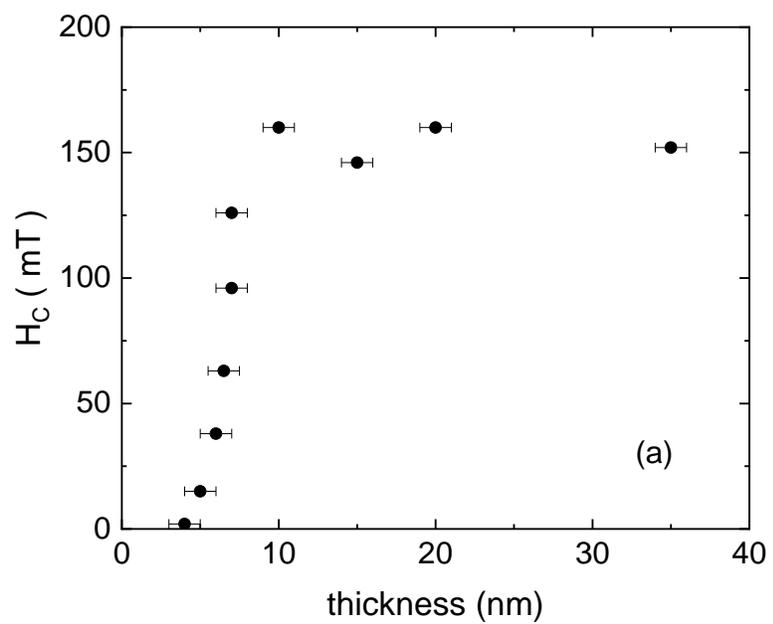

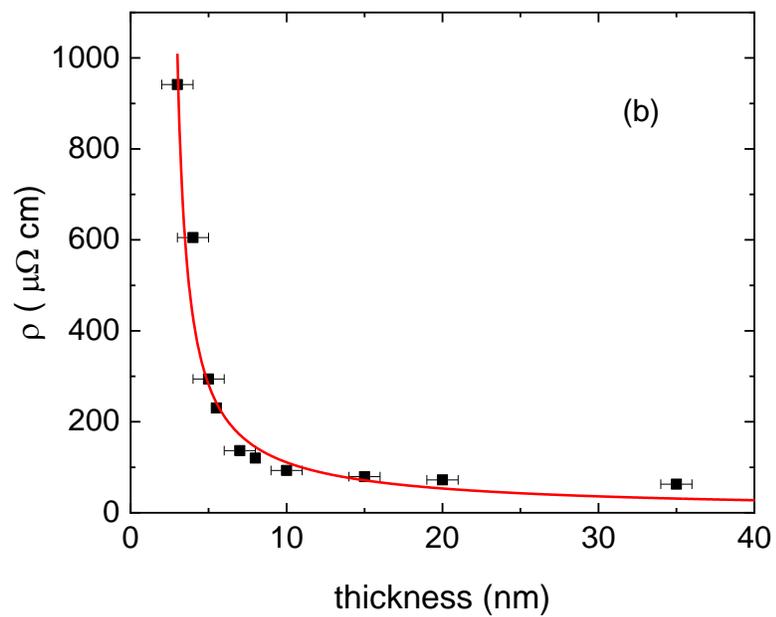

Fig. 2

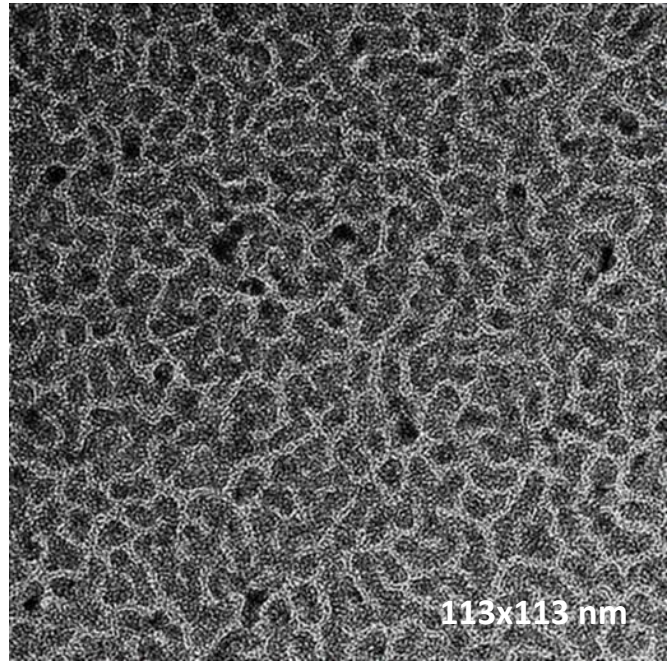

Fig. 3

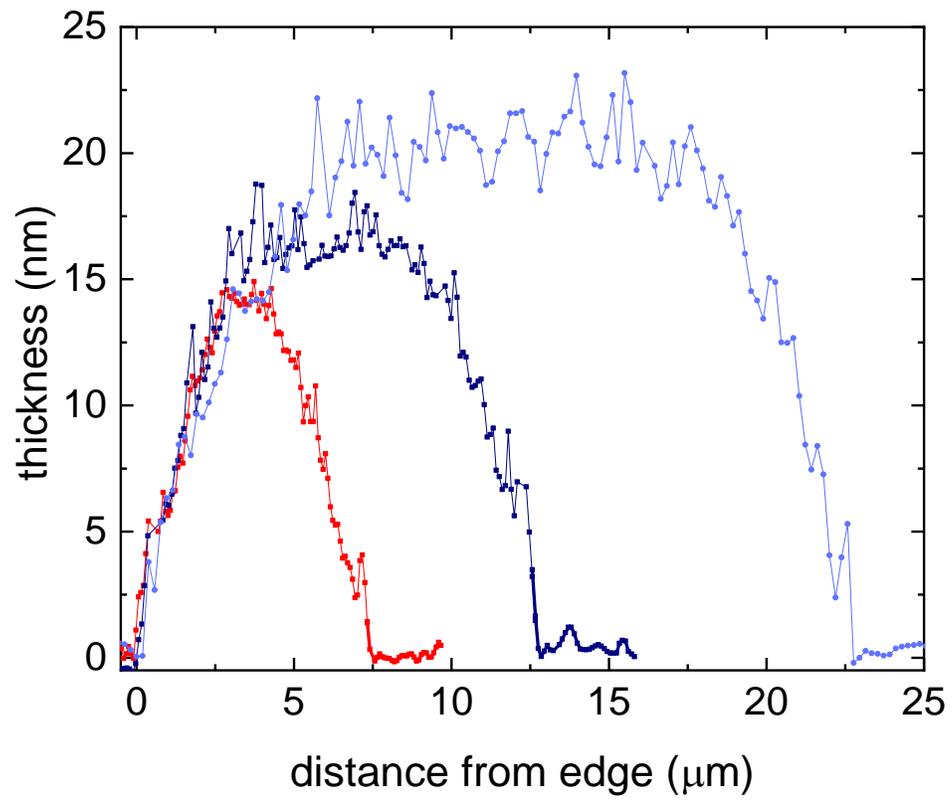

Fig. 4

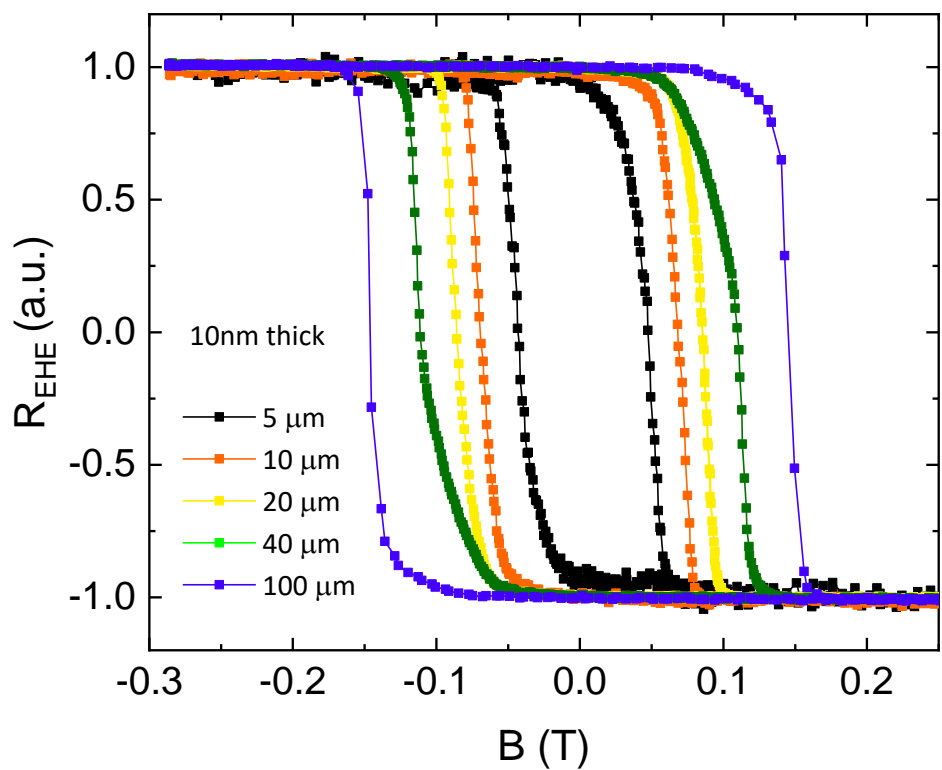

Fig. 5

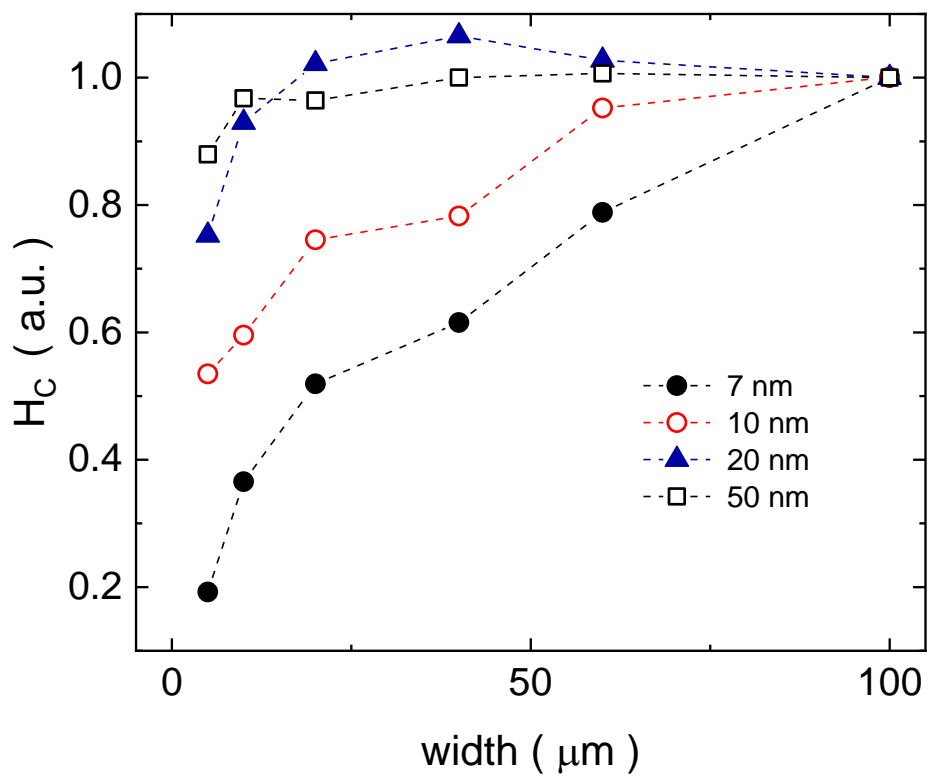

Fig. 6

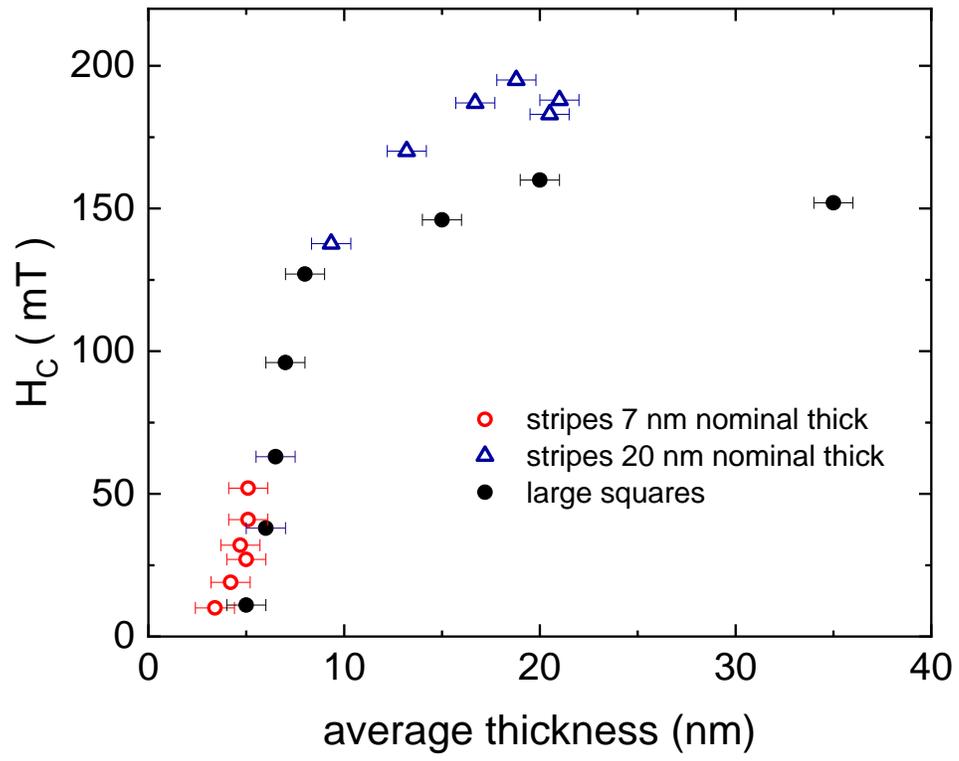

Fig. 7